\documentclass[showpacs,preprint,aps]{revtex4}

\usepackage{graphicx}
\usepackage{dcolumn}
\usepackage{bm}
\usepackage{color}

\newcommand\E{\varepsilon}
\def\vec{\mathbf}

\begin{document}

\preprint{APS/123-QED}

\title{Resonances in small scatterers with impedance boundary}

\author{Ari Sihvola, Dimitrios C. Tzarouchis, Pasi Yl\"a-Oijala,  
Henrik Wall\'en, and Beibei Kong}
\affiliation{%
Aalto University School of Electrical Engineering\\
Department of Electronics and Nanoengineering\\
Espoo, Finland
}%


\date{\today}

\begin{abstract}
  With analytical (generalized Mie scattering) and numerical
  (integral-equation-based) considerations we show the existence of
  strong resonances in the scattering response of small spheres with
  lossless impedance boundary. With increasing size, these multipolar
  resonances are damped and shifted with respect to the magnitude of
  the surface impedance. The electric-type resonances are inductive
  and magnetic ones capacitive. Interestingly, these subwavelength
  resonances resemble plasmonic resonances in small
  negative-permittivity scatterers and dielectric resonances in small
  high-permittivity scatterers. The fundamental dipolar mode is also
  analyzed from the point of view of surface currents and the effect
  of the change of the shape into a non-spherical geometry.
\end{abstract}

\maketitle

Slight perturbations in a system lead usually to small changes in its
response function. In electromagnetic scattering, a good example is
Rayleigh scattering which means that the total scattered power of a
particle small compared with the wavelength is proportional to the
sixth power of its linear size, and thus vanishes predictably for tiny
objects. But there are exceptions: a plasmonic subwavelength particle
may have a very strong scattering response. This happens, for example,
when the relative permittivity of a sphere hits the value $-2$, and
the particle is capable of supporting a localized dipolar-like surface
plasmon. The magnitude of the resonance is in practice attenuated by
material losses and radiation damping but can in principle reach very
large values for spheres with diameter much smaller than the
wavelength of the incident radiation. In this letter we report on a
similar phenomenon that can be found in other type of small
scatterers: our analysis shows that particles with a special type of
impedance boundary can have scattering and extinction efficiencies
that grow without limit when their size decreases in the subwavelength
domain. This phenomenon may have fundamental implications regarding
the scattering by optically small objects.

In electromagnetics, a multitude of boundary condition exists that can
be classified into different categories \cite{Lindell_GenDB}. Among
those, much-used are the PEC (perfect electric conductor) and PMC
(perfect magnetic condutor) boundaries, on which the tangential
electric (PEC) or tangential magnetic (PMC) field has to vanish.
These conditions are special cases of the impedance boundary
condition (IBC) which requires the following relation between the
tangential electric $(\vec{E}_t)$ and magnetic $(\vec{H}_t)$ fields:
\begin{equation}
  \label{eq:ibc}
  \vec{E}_t = Z_s  \vec{n}\times \left( \eta_0\vec{H}_t\right)
\end{equation}
on the surface with unit normal vector $\vec n$. The surface impedance
is a naked number, having units of free-space impedance
$\eta_0=\sqrt{\mu_0\E_0}$. For the impedance surface to be lossless,
$Z_s$ has to be purely imaginary \cite[Section 3.6]{methods}. A
passive (dissipative) surface has positive real part of $Z_s$, and
correspondingly the negative real part means an active surface. 

The history of the IBC concept reaches back to 1940's
\cite{Shchukin,Leontovich} when it was introduced in connection with
the analysis of radio-wave propagation over ground.  The scattering
problem involving IBC objects has been treated in some studies in the
past \cite[Section 10.4]{Jackson}, \cite{Glisson} but it seems that
the fundamental phenomenon of resonance modes in small particles has
not received earlier attention. The understanding of such mechanisms
in the scattering problem opens up possibilities to tailor structures
with desired electromagnetic response. This complements other
approaches that exist to engineer the scattering characteristics of
material objects, such as metasurface-based manipulations
\cite{ele,bilo,glybo} and various principles to reduce visibility,
like mantle cloaking \cite{Alu}.


Let us first compute the interaction of an IBC sphere with incident
electromagnetic plane wave in free space. The incident field will be
scattered from the sphere, and the scattered field can be expanded in
an infinite series of spherical harmonic functions. The expansion
coefficients follow from the boundary condition at the impenetrable
surface of the sphere. Following {\em mutatis mutandis} the classical
Lorenz--Mie analysis, we arrive at the electric ($a_n$) and magnetic
($b_n$) scattering coefficients

\begin{eqnarray}
  \label{eq:a1}
  a_n & = & \frac{x\, \mathrm{j}_{n-1}(x)-n\, \mathrm{j}_n(x)
+ \mathrm{i} Z_s \,x\,\mathrm{j}_n(x)}
  {x\, \mathrm{h}_{n-1}^{(1)}(x)-n \,\mathrm{h}_n^{(1)}(x)
+ \mathrm{i} Z_s \, x \,\mathrm{h}_n^{(1)}(x)}  \\
  \label{eq:b1} 
 b_n & = & \frac{x \,\mathrm{j}_{n-1}(x)-n \,\mathrm{j}_n(x) + 
                 (\mathrm{i}/Z_s)\, x \,\mathrm{j}_n(x)}
  {x \,\mathrm{h}_{n-1}^{(1)}(x)-n \,\mathrm{h}_n^{(1)}(x) +
                   (\mathrm{i}/Z_s)\, x \,\mathrm{h}_n^{(1)}(x)}
\end{eqnarray}
Here $x=2\pi a/\lambda$ is the size parameter of the sphere with
radius $a$, and $\mathrm{j}_n$ and $\mathrm{h}_n^{(1)}$ are the
spherical Bessel and Hankel (of the first kind) functions of order
$n$. The convention $\exp(-\mathrm{i}\omega t)$ is applied to map the
sinusoidal time-dependence into complex numbers. From these
coefficients, the scattering (sca), extinction (ext), and absorption
(abs) efficiencies can computed according to the same principle as
with classical efficiencies of penetrable spheres \cite[Section
4.4]{Bohren_Huffman}:
\begin{eqnarray}
  \label{eq:Qsca}
  Q_\mathrm{sca} & = & \frac{2}{x^2} \sum_{n=1}^\infty (2n+1) \left( |a_n|^2 + |b_n|^2 \right) \\
  Q_\mathrm{ext} & = &  \label{eq:Qext}
\frac{2}{x^2} \sum_{n=1}^\infty (2n+1) \mathrm{Re}\left\{ a_n + b_n \right\} \\
  Q_\mathrm{abs} & = & Q_\mathrm{ext} - Q_\mathrm{sca} \label{eq:Qabs}
\end{eqnarray}
The efficiency is a dimensionless figure of merit, {\em e.g.}, the
scattering efficiency is the scattering cross section divided by the
geometrical cross section of the particle. The series (\ref{eq:Qsca})
and (\ref{eq:Qext}) converge. The larger the sphere in terms of
wavelength, the more terms are needed. We use the Wiscombe criterion
for the necessary number of terms ($N_\mathrm{max}=x+4\sqrt[3]{x}+2$)
to truncate the series \cite{Wiscombe}.

\begin{figure}
\null\vspace{5mm}
\includegraphics[width=8cm]{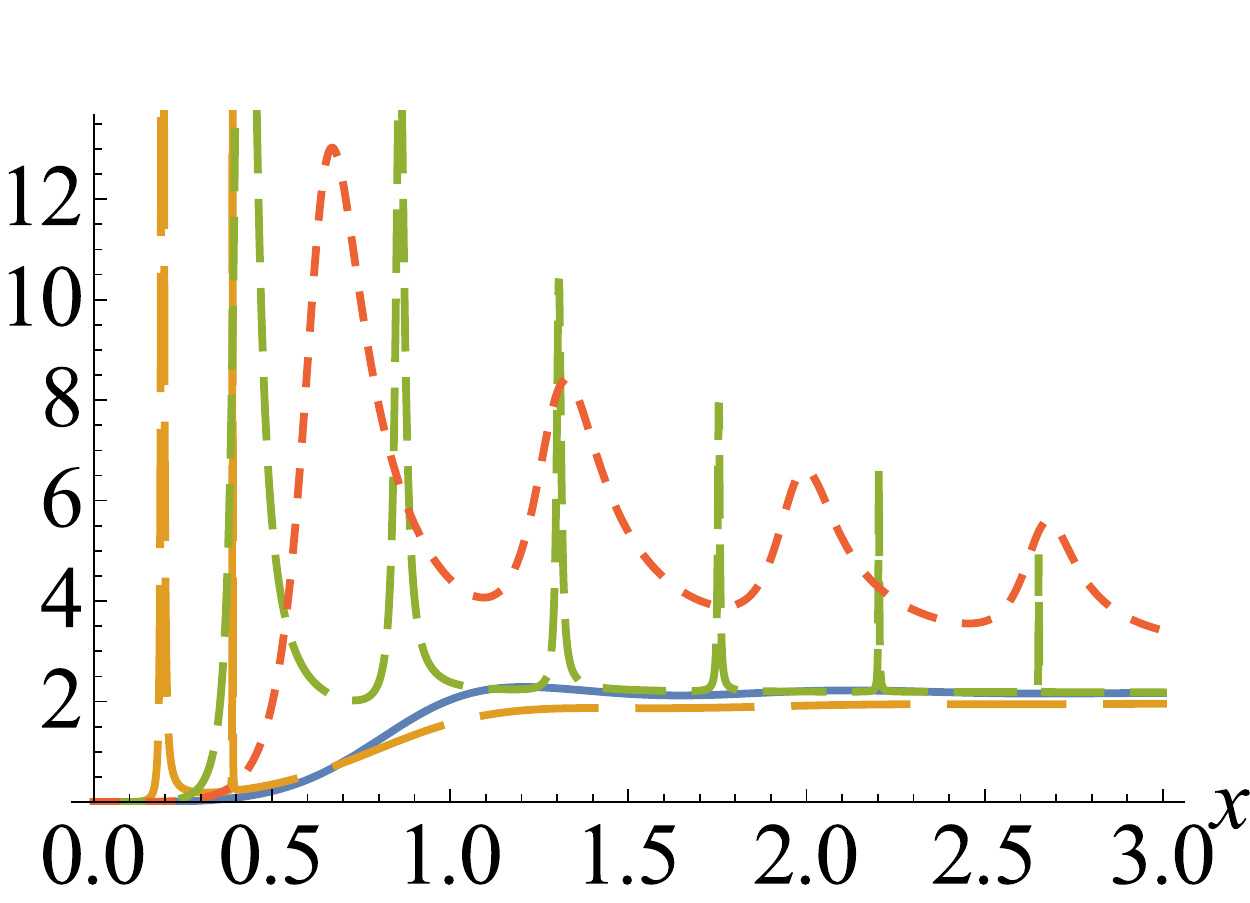} \hspace{1mm}
\includegraphics[width=8cm]{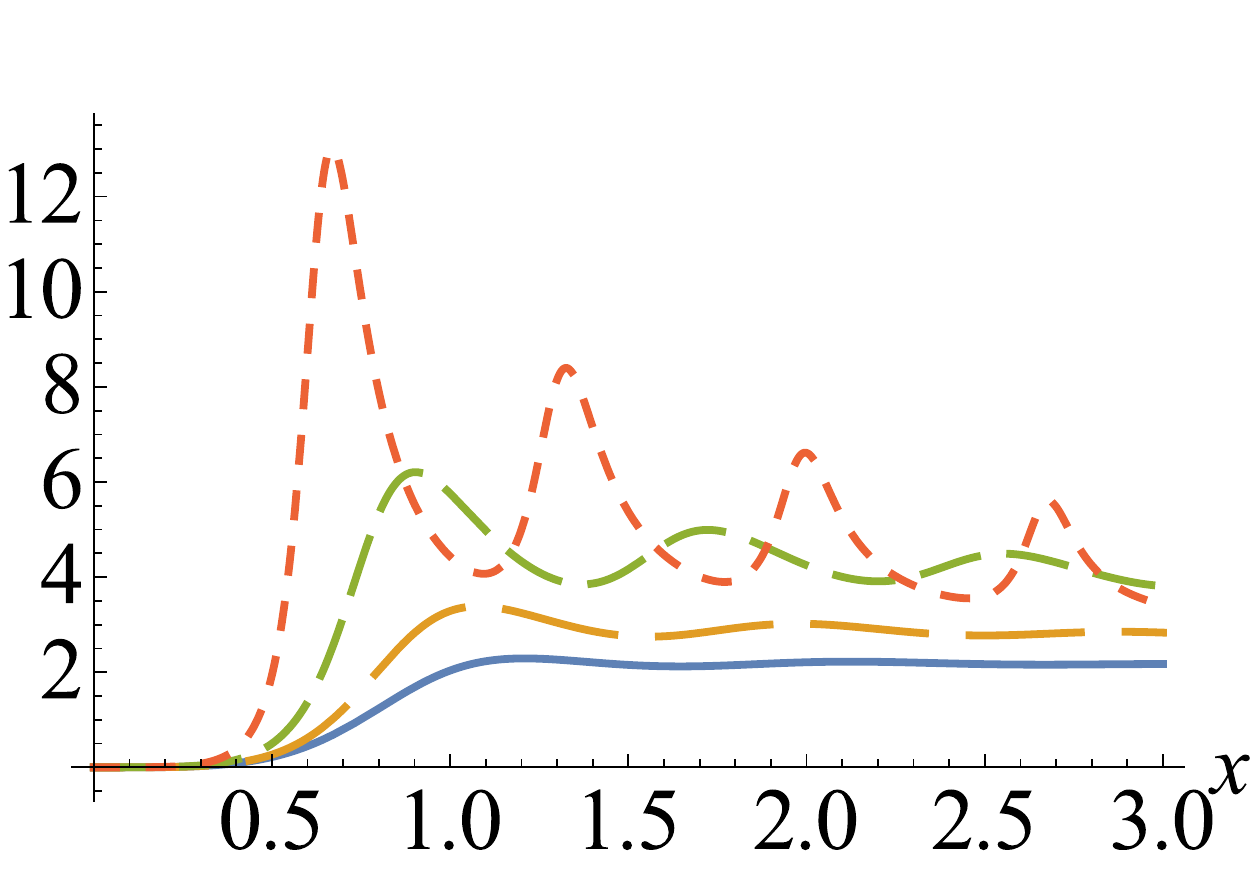}
\caption{\label{fig:Q2} Scattering (extinction) efficiency of lossless
  impedance spheres as functions of size parameter $x$, for certain
  values of $X_s$ which is the (negative of the) normalized surface
  reactance: (left) solid blue: $X_s=0$ (PEC), long-dashed orange: $X_s=-0.2$,
  short-dashed green: $X_s=-0.5$, dotted red: $X_s=-1$; (right): solid
  blue: $X_s=\pm\infty$ (PMC), long-dashed orange: $X_s=-5$,
  short-dashed green: $X_s=-2$, dotted red: $X_s=-1$.}
\end{figure}

With this mathematical equipment, we can calculate the scattering and
extinction behavior of spheres with arbitrary surface impedance and
size. For lossy scatterers ($Z_s$ has a non-zero real part), all three
efficiencies are different while in the lossless case ($Z_s$ is purely
imaginary), the absorption efficiency vanishes and
$Q_\mathrm{sca}=Q_\mathrm{ext}$. Following the notation of circuit
theory, we write the surface impedance as
\begin{equation}
  Z_s = R_s - \mathrm{i}X_s
\end{equation}
into the surface resistance $R_s$ and surface reactance $X_s$. The
reactance $X_s$ is positive for inductive surfaces, and negative for
capacitive. The particularly interesting finding from our studies
concerns lossless scatterers for which the surface impedance is
$Z_s=-\mathrm{i}X_s$. We plot the scattering efficiencies of IBC
spheres as functions of size parameter $x$ for different values of the
surface impedance in Figure~\ref{fig:Q2}. Due to the lossless
character of the sphere, the scattering efficiency equals the
extinction efficiency.

As to their scattering efficiency, the PEC ($Z_s=0$) and PMC
($1/Z_s=0$) spheres behave identically (the two blue curves in
Figure~\ref{fig:Q2}). However, the functional form of the responses is
far from trivial for intermediate surface impedance values. As the
value of the reactance $X_s$ decreases from the PMC limit, a gradual
increase in scattering amplitude takes place over all the range. The
evolution leads to oscillations, and once the surface reactance
reaches small (negative) values, the whole curve is dominated by the
resonances. In the PEC limit ($X_s$ is close to zero), the resonances,
riding on top of the gently rolling PEC curve, become vanishingly
narrow.

\begin{figure}
\null\vspace{5mm}
\includegraphics[width=8cm]{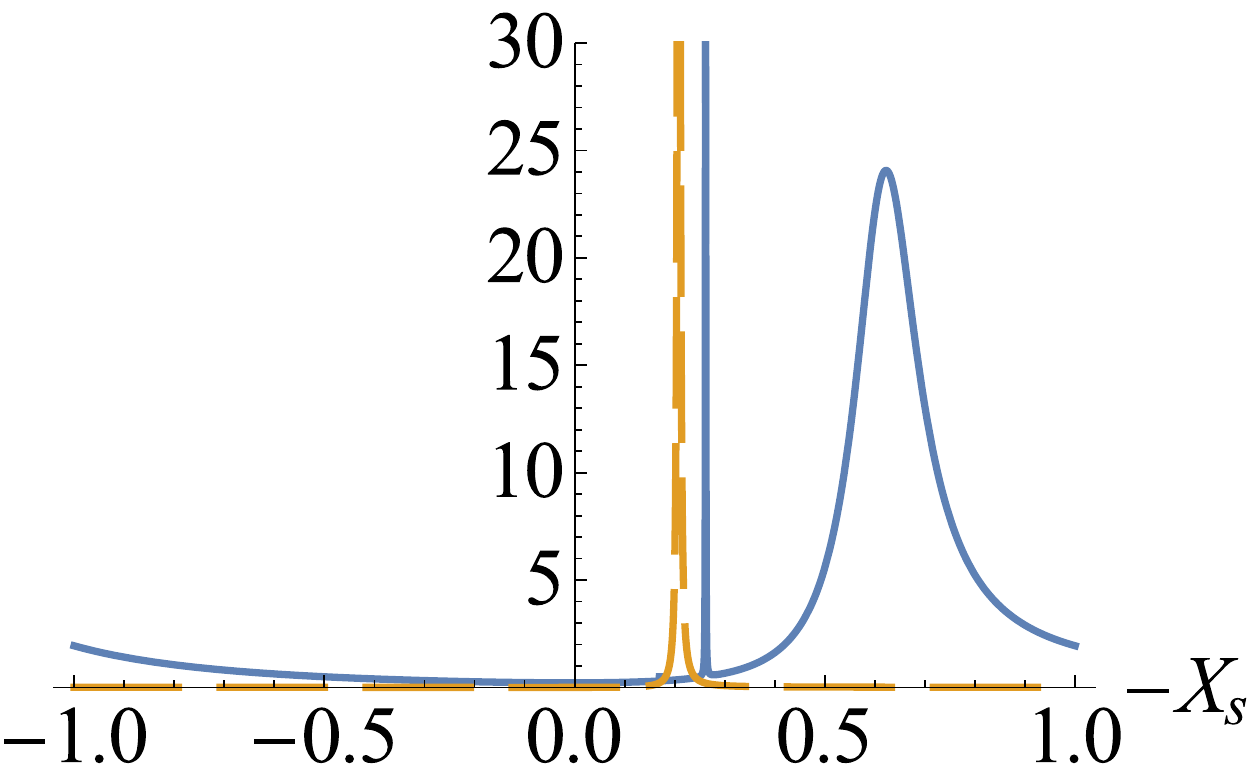} \hspace{1mm}
\includegraphics[width=8cm]{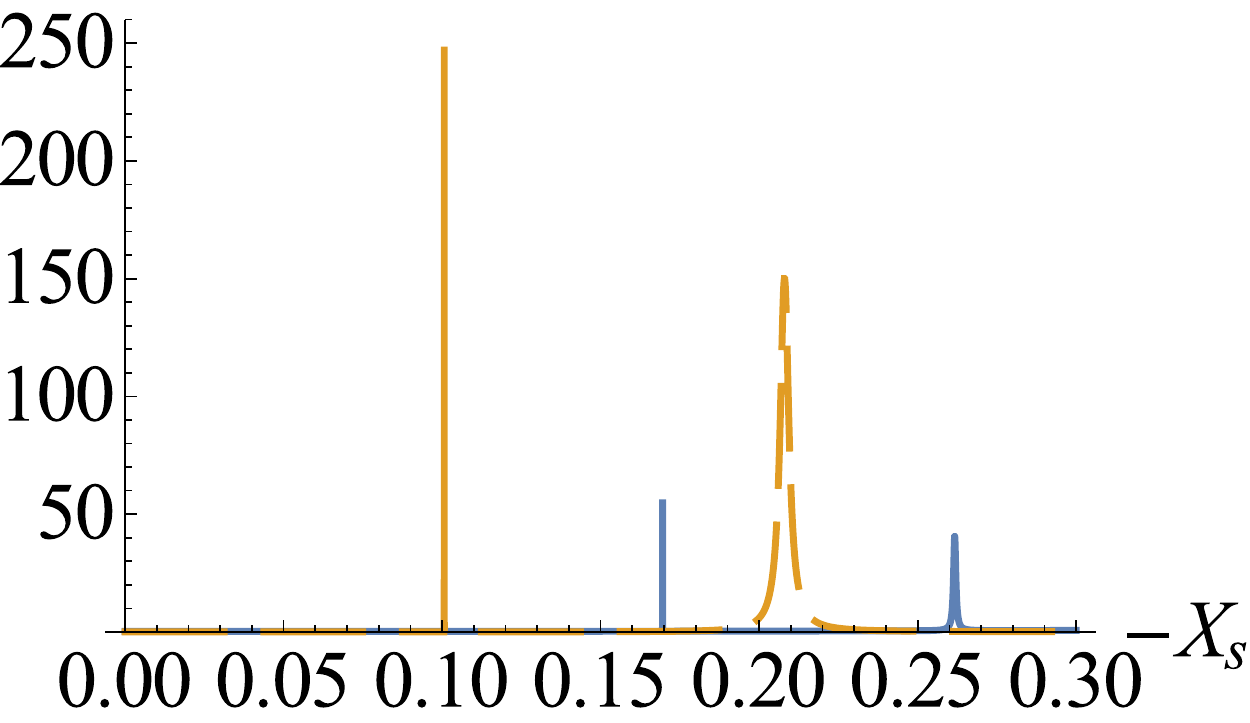}
\caption{\label{fig:QQ2} Scattering (extinction) efficiency of
  lossless impedance spheres as functions of the surface reactance
  parameter $X_s$.  The size parameter is $x=0.5$ (solid blue) and
  $x=0.2$ (long-dashed orange). Due to the broad range of $X_s$ in the
  left-hand side figure, very high-order resonances cannot be
  distinguished.}
\end{figure}

The broadest (dipolar) resonance for very small spheres appears at
$-X_s\approx x$, and the higher-order modes follow with $-X_s\approx
x/n$ for integers $n$. Figure~\ref{fig:QQ2} shows a closer view the
resonances, showing the positions of quadrupolar and octopolar modes
for size parameters $x=0.5$ and $x=0.2$. The amplitude of the
resonances is $\mathrm{max}\{Q_\mathrm{sca}\} \approx 2(2n+1)x^{-2}$, being
$24,40,64$ for the three lowest modes in case of $x=0.5$ and
$150,250,350$ for $x=0.2$.

\begin{figure}
\null\vspace{5mm}
\includegraphics[width=12cm]{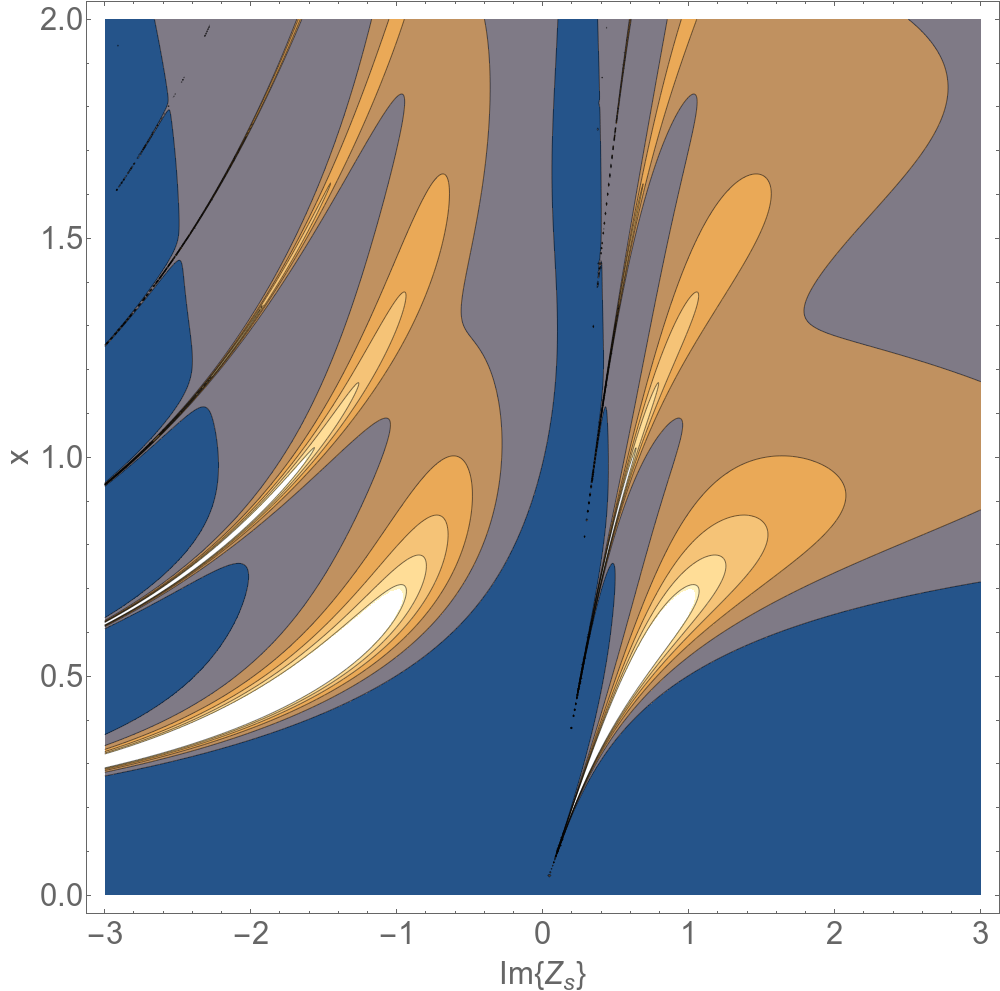}
\caption{\label{fig:Q3D} A contour plot of the scattering efficiency
  of an impedance sphere as function of its surface reactance and the
  size parameter $x$.}
\end{figure}

Figure~\ref{fig:Q3D} illustrates the scattering characteristics as a
function of the size parameter and the surface impedance. The effect
of increasing sphere size is to soften the resonances and shift their
position towards larger values of the imaginary part of the surface
impedance. Two clusters of resonance modes exist, one for positive and
one for negative $X_s$. The resonances at $\mathrm{Im}\{Z_s\}>0$ are
due to the maxima of the $b_n$ Mie coefficients (\ref{eq:b1}), and
hence are magnetic type resonances, while the resonances for negative
$\mathrm{Im}\{Z_s\}$ arise from $a_n$ (\ref{eq:a1}), being of electric
type. Despite the different visual appearance of the two clusters,
they follow the symmetry
\begin{equation}\label{eq:symm}
  Q_\mathrm{sca}(x,X_s) =  Q_\mathrm{sca}(x,-1/X_s)
\end{equation}

\begin{figure}[h!]
\null\vspace{5mm}
\includegraphics[width=8cm]{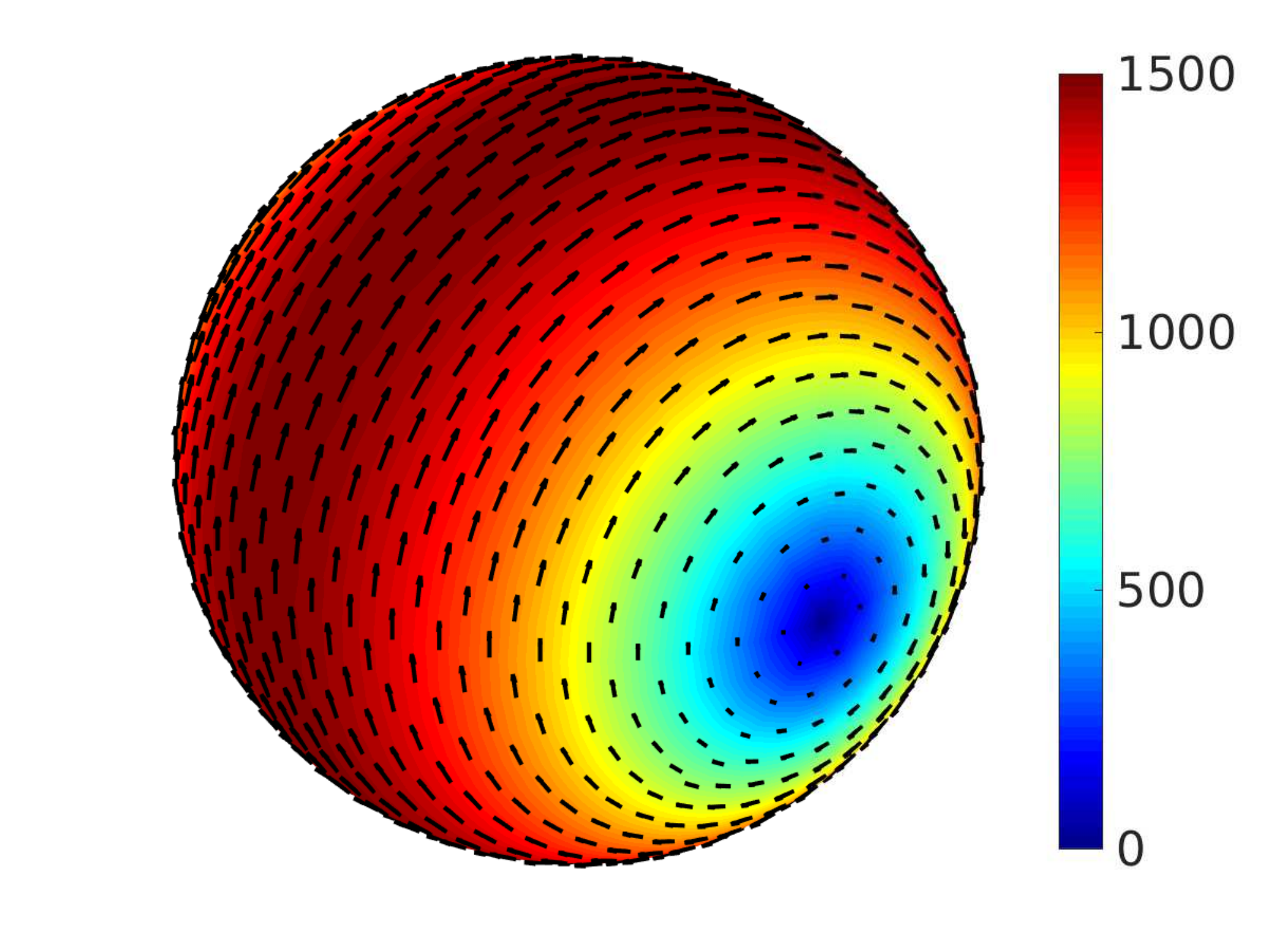} \hspace{1mm}
\includegraphics[width=8cm]{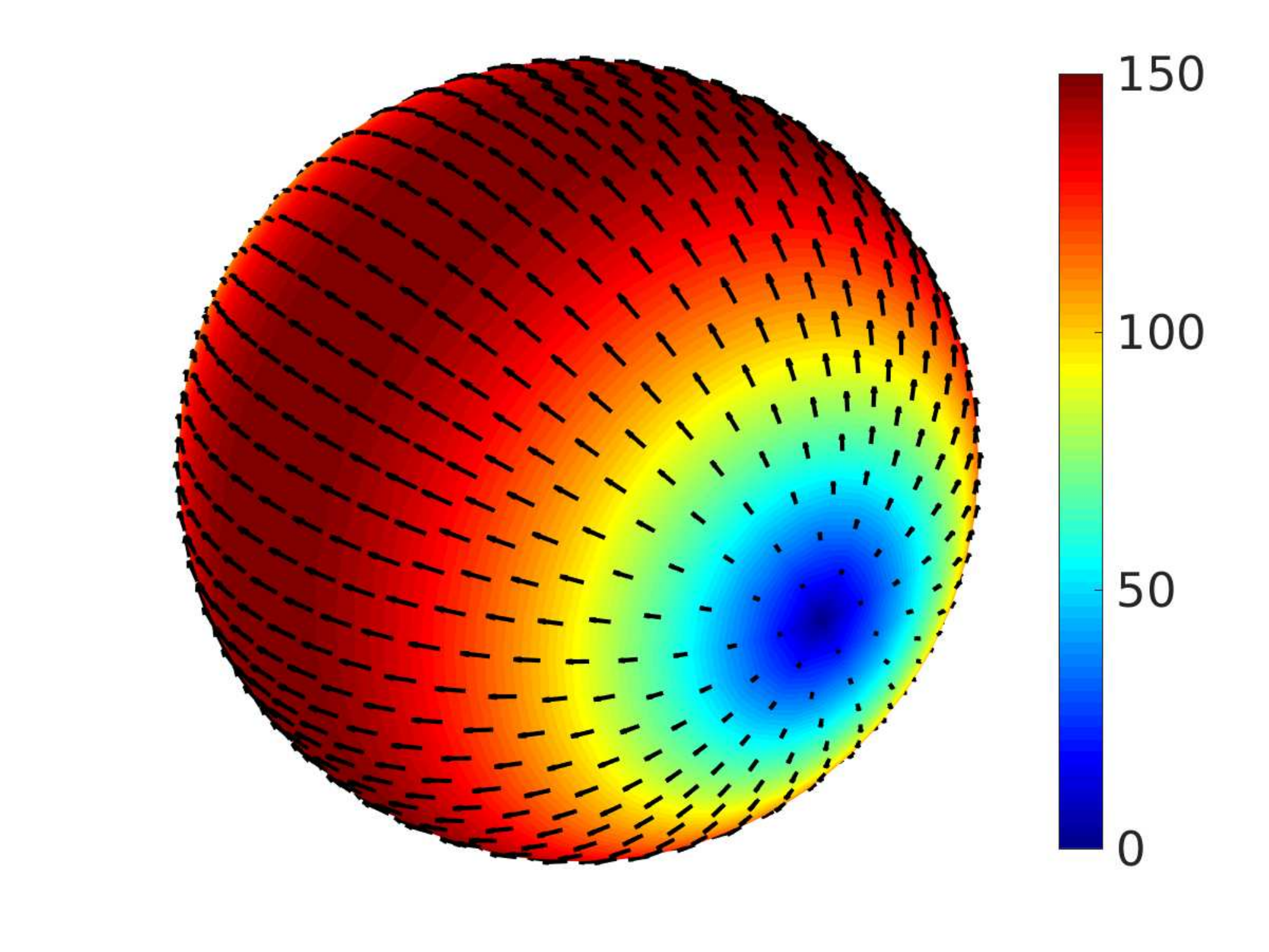} 
\caption{\label{fig:realM} The imaginary part of the electric surface
  current (left) and the real part of the magnetic surface current
  (right) for an IBC sphere with size parameter $x=0.1$ and
  homogeneous scalar surface impedance $Z_s=\mathrm{i}\,0.101$. The
  incident wave is propagating upwards from the bottom.}
\end{figure}

To gain understanding of the mode pattern of the lowest resonance,
Figure~\ref{fig:realM} displays the induced electric $\vec{J}_s$ and
magnetic $\vec{M}_s$ surface currents on the sphere for the case
$x=0.1$ and $Z_s=\mathrm{i}\,0.101$, with a plane wave excitation. The
surface currents are connected to the tangential fields as $\vec{J}_s
= \vec{n}\times\vec{H}$ and $\vec{M}_s = -\vec{n}\times\vec{E}$. The
current distributions show clearly a magnetic dipole type of structure
(circulating electric current). Due to the boundary condition
(\ref{eq:ibc}) where the surface impedance is imaginary, the currents
have to be in $90^\circ$ phase shifted, and also rotated by $90^\circ$
on the sphere surface. The tenfold magnitude of the electric current
compared with the magnetic follows from amplitude of the surface
reactance. (The figures display only the imaginary(real) part of
$\vec{J}_s$ $(\vec{M}_s)$; the out-of-phase components are around 3000
times smaller.)

\begin{figure}
\null\vspace{5mm}
\includegraphics[width=8cm]{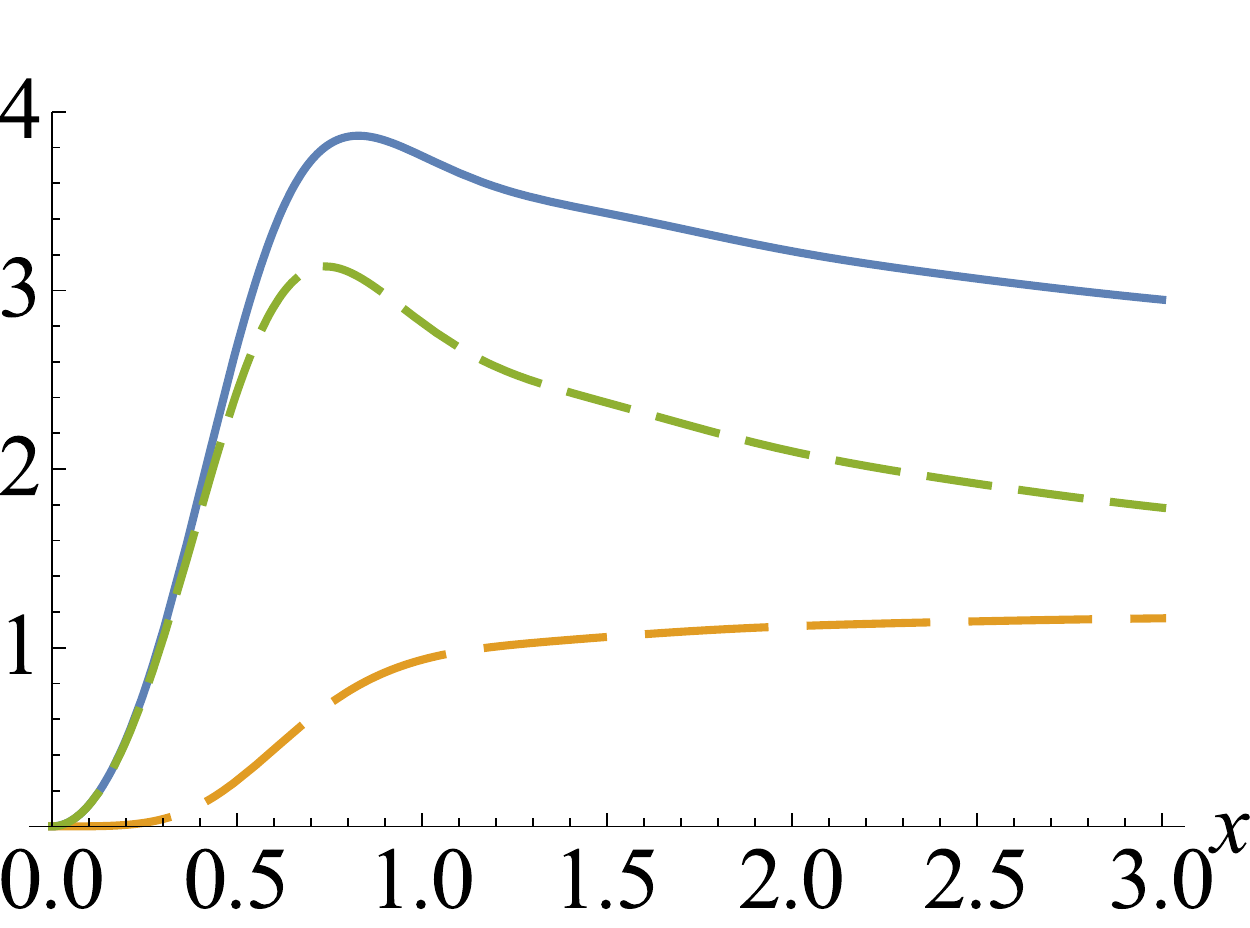} \hspace{1mm}
\includegraphics[width=8cm]{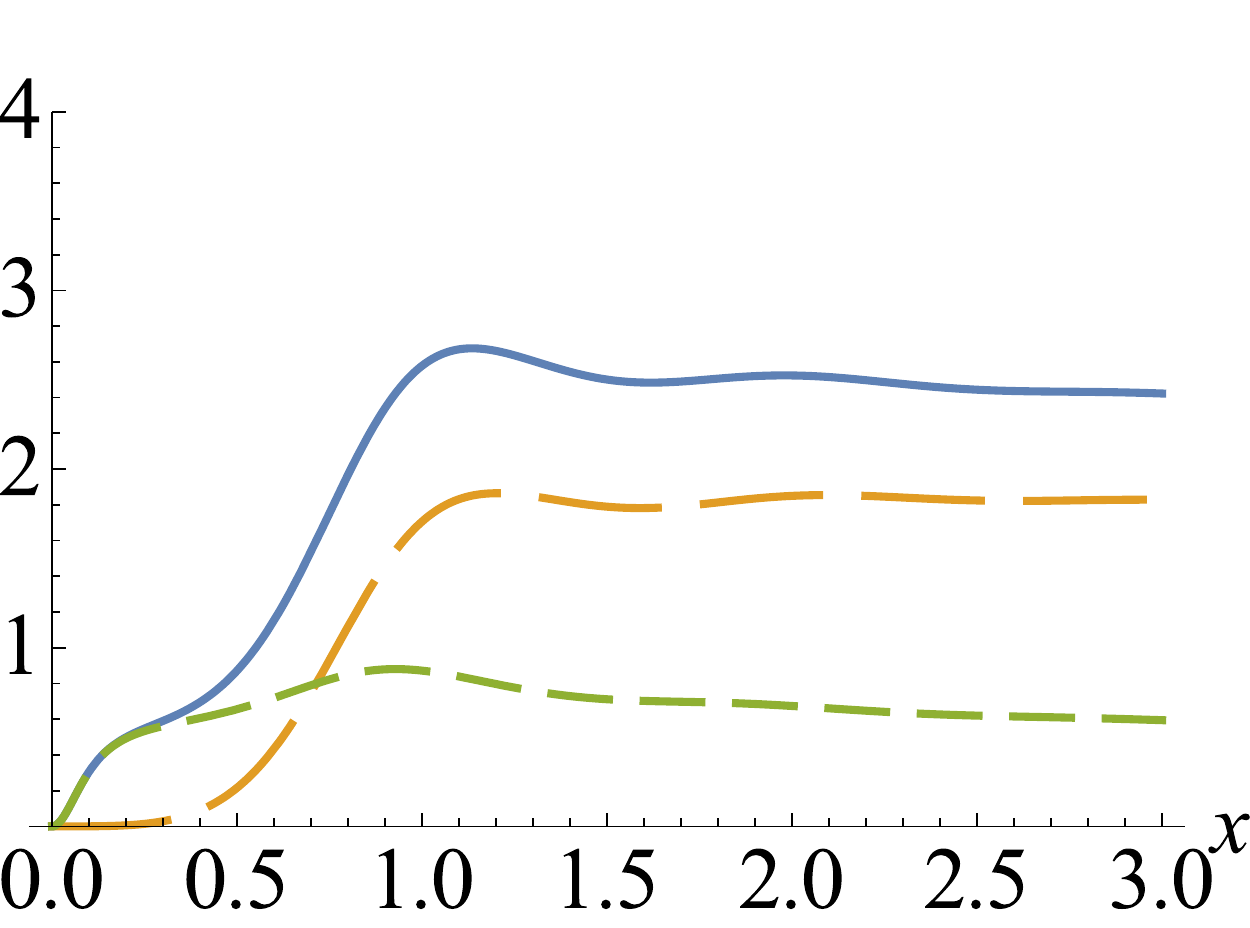}
\caption{\label{fig:loss} Extinction (solid blue), scattering
  (long-dashed orange), and absorption (short-dashed green)
  efficiencies for spheres with lossy surface impedance as a function
  of the size parameter: $Z_s=1$ (left), $Z_s=10$ (right).}
\end{figure}

When the surface impedance $Z_s$ has a real part, the surface is no
longer lossless. Hence also the three efficiencies in
(\ref{eq:Qsca})--(\ref{eq:Qabs}) are different. For passive surfaces
(real part of $Z_s$ is positive), there is absorption
$(Q_\mathrm{abs}>0)$, and in case of active surfaces, absorption is
negative. The interplay between scattering, absorption, and extinction
for lossy impedance spheres is depicted in Figure~\ref{fig:loss} for
two cases: $Z_s$ has real and positive values $1$ and $10$.

\begin{figure*}[h!]
\null\vspace{5mm}
\includegraphics[height=4.8cm]{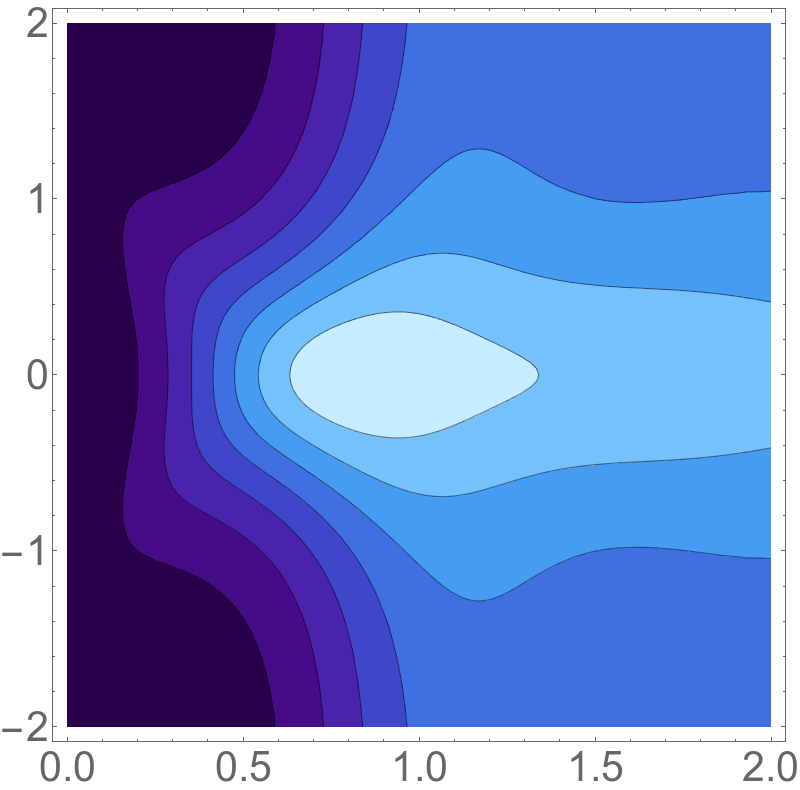} \hspace{0mm}
\includegraphics[height=4.8cm]{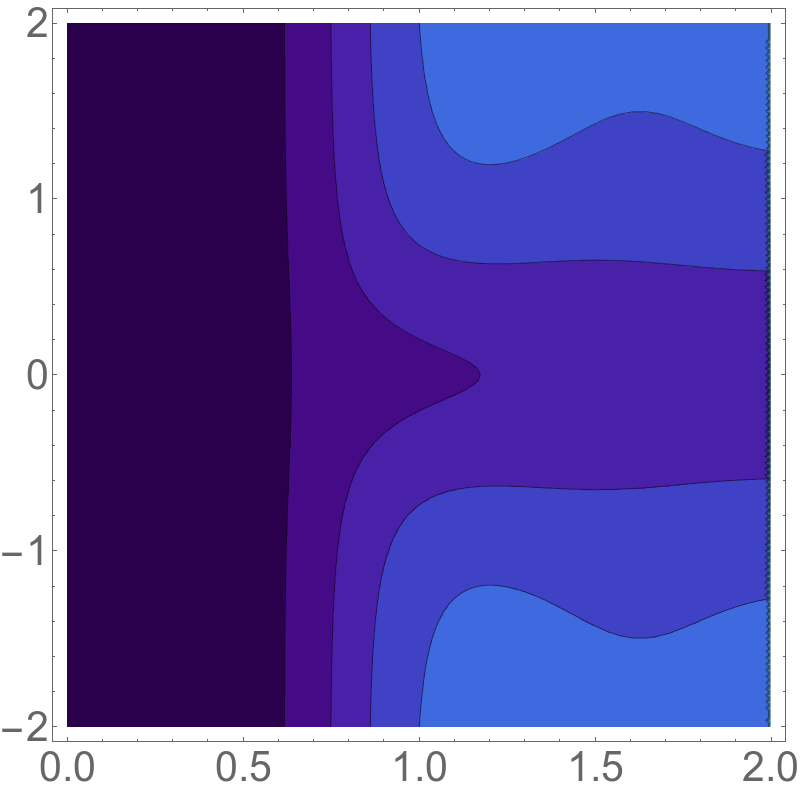} \hspace{0mm}
\includegraphics[height=4.8cm]{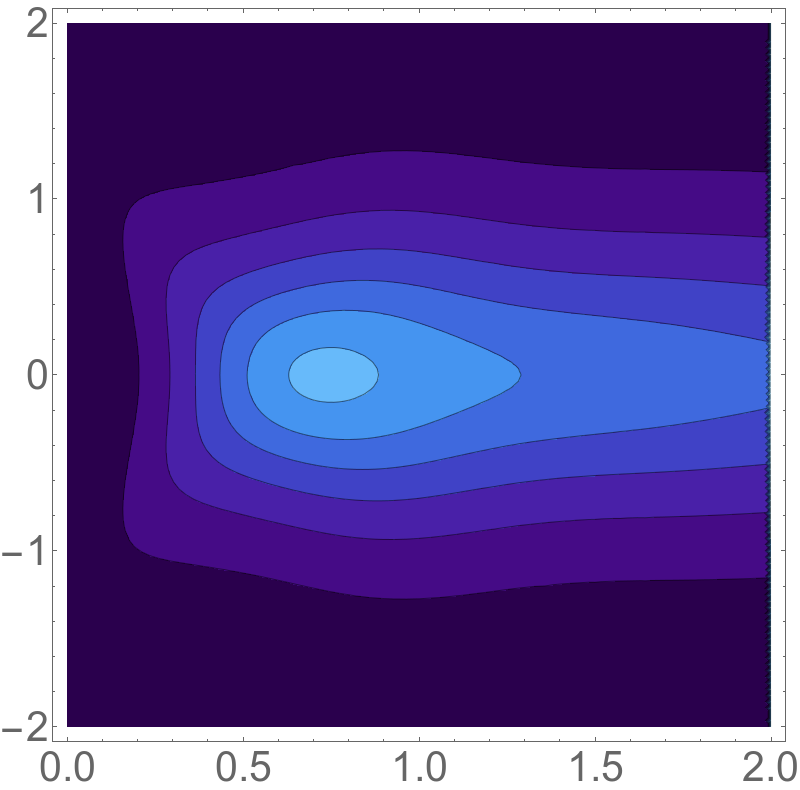} \hspace{0mm}
\includegraphics[height=3.3cm]{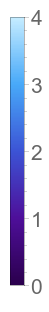}
\caption{\label{fig:w} Extinction (left), scattering (center), and
  absorption (right) efficiencies of a lossy IBC sphere as functions
  of the size parameter (horizontal axis $x$) and the (logarithmic)
  surface impedance (vertical axis $a$ with $Z_s=10^a$).}
\end{figure*}

The dominance of absorption over scattering is conspicuous for small
spheres in Figure~\ref{fig:loss}: the extinction is mainly due to
absorption when $x$ is small. This phenomenon is also seen in the
Taylor expansions of the scattering efficiencies (\ref{eq:Qsca}) and
(\ref{eq:Qext}). The ordinary Rayleigh scattering dependence
$Q_\mathrm{sca}\approx (16/3) x^4$ holds for scattering, while the
absorption efficiency has a square dependence on the size parameter:
\begin{equation}
Q_\mathrm{abs}\approx 6(Z_s+1/Z_s)x^2  
\end{equation}
This square dependence of absorption efficiency on size parameter of
IBC spheres differs from the corresponding behavior of the absorption
of lossy penetrable dielectric spheres in which the dependence is
linear in the small-particle limit \cite{rs}. For small spheres with
fixed $x$, the maximum absorption takes place for increasing $Z_s$ or
$1/Z_s$, but as the size parameter is larger than $0.4$, the maximum
is strongest for the ``matched-impedance'' case $Z_s=1$ (for which the
scattering achieves its minimum). Such a sphere is an example of a
zero-backscattering object \cite{ZBO}.

The efficiencies are invariant with respect to the change
$Z_s\rightarrow 1/Z_s$. Therefore (\ref{eq:symm}) can be written for
general complex IBC spheres as $Q(x,Z_s)=Q(x,1/Z_s)$, valid for all
three efficiency quantities and $Z_s=R_s-\mathrm{i}X_s$.

Figure~\ref{fig:w} shows contour plots of the three efficiencies as
functions of the size parameter and the (real-valued) surface
impedance. In agreement with the extinction paradox (\cite[Section
4.4.3]{Bohren_Huffman}, \cite{Brillouin}), the extinction efficiency
approaches the value $2$ for large spheres, independent of the surface
impedance. The convergence can be slow if $|Z_s|\approx 1$: to be
within one percent of this limiting value, $x$ needs to be around one
thousand.

\begin{figure}[h!]
\null\vspace{5mm}
\includegraphics[width=12cm]{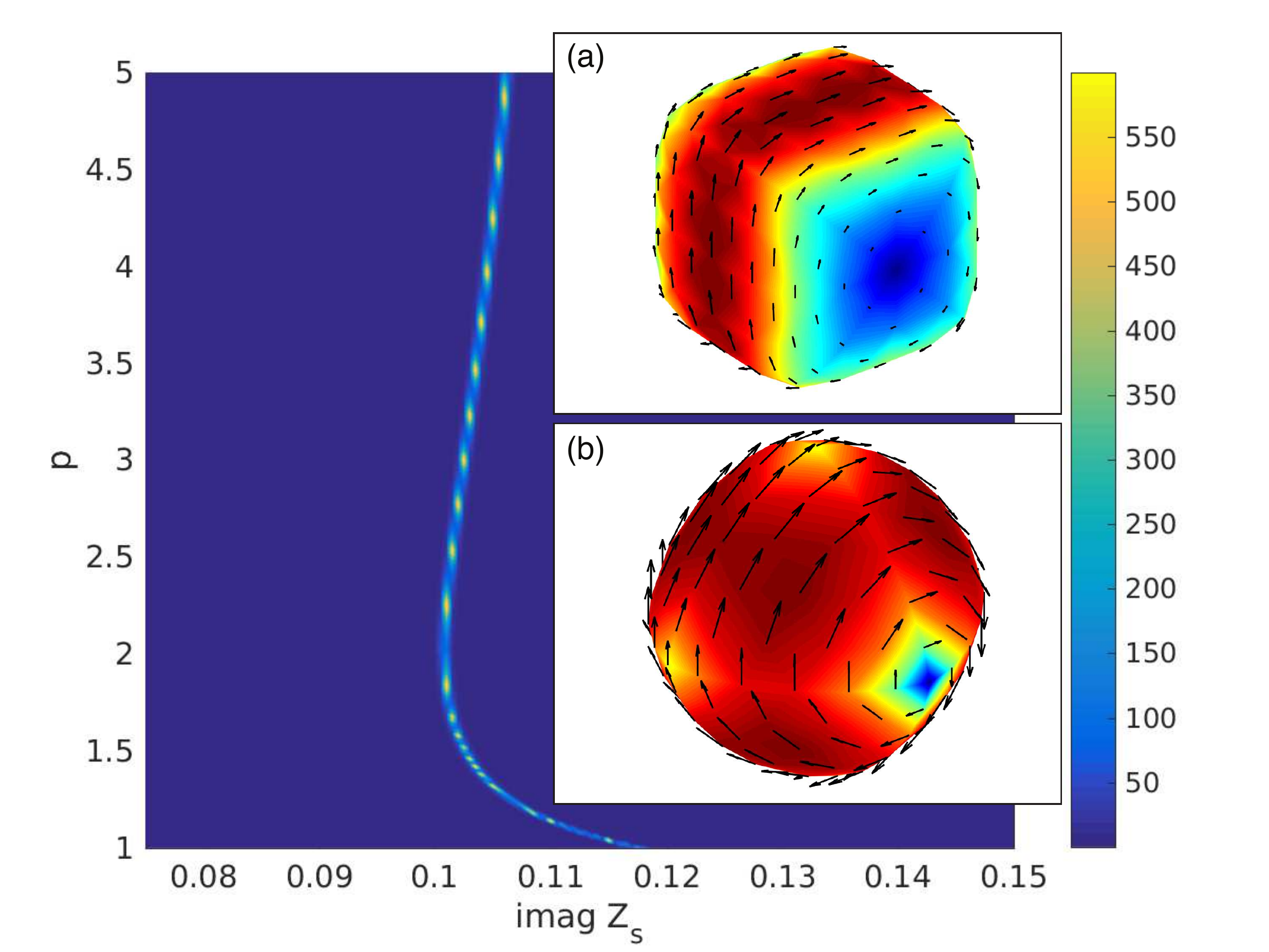} 
\caption{\label{fig:Zsp} The magnitude of the scattering efficiency of
  a small lossless IBC quasispherical particle in the plane of the
  surface impedance and the supersphere parameter $p$. Throughout the
  $p$ scale, the volume of the particle is the same as the volume of a
  sphere with $p=2$ for which $x=0.1$. The insets show the electric
  surface current distributions for cases (a) $p=5$ and (b) $p=1.2$.}
\end{figure}

Sphere is an extravagantly symmetric shape. It is fair to raise the
question whether the IBC resonances remain if the spherical symmetry
is broken. As an answer we compute the scattering behavior in the
vicinity of the resonance of the magnetic dipole for an IBC
superspherical object, using numerical surface-integral-equation
method based on the electric field integral equation formulation for
IBC scatterers \cite{eibert,pasi}. The surface of such an object is
defined by
\begin{equation}
  |x|^p + |y|^p +|z|^p = a^p
\end{equation}
The value $p=2$ reproduces a sphere with radius $a$, $p=1$ an
octahedron, and for increasing $p$, the shape approaches a
cube \cite{Dimi_josa}. Figure~\ref{fig:Zsp} shows how the scattering response, in
particular how the position of the main resonance shifts with the
shape of the object. Not surprisingly, the spherical geometry gives an
extremum.

The question about material realization of these scatterers
remains. In engineering electromagnetics, the boundary condition has
been used as an approximation to interfaces between materially
strongly contrasting media, often with success, like in the response
of metal surfaces in the microwave region. However, interfaces over
which the material parameters only change moderately cannot be
accurately described by a boundary condition due to the fact that the
relation between the tangential electric and magnetic fields depends
on the incidence angle and polarization of the incident wave. A remedy
to the synthesis problem is the so-called wave-guiding material
\cite{Lindell06a} which is an anisotropic medium that has large
components in the permittivity and permeability dyadics in one
direction. Cut perpendicular to this direction, the material surface
behaves like an impedance surface since the large values of the
normally directed material parameter components force the longitudinal
fields to vanish, and the fields in the material remain
transversal. Adding a parallel metallic plate at a certain depth, the
waves are reflected and travel like in a waveguide, and the field
relation can be manipulated by varying the transversal permittivity
and permeability components of the medium. Another approach to
materialize a structure mimicking the surface reactance is to make use
of use of frequency-selective-surface (FSS) principles \cite{Munk} and
carve a regular subwavelength pattern of holes on a conducting
metallic surface, thus manipulating the ratio between the averaged
electric and magnetic fields to produce the desired surface impedance.


%


\begin{thebibliography}{1}%
\makeatletter
\providecommand \@ifxundefined [1]{%
 \ifx #1\undefined \expandafter \@firstoftwo
 \else \expandafter \@secondoftwo
\fi
}%
\providecommand \@ifnum [1]{%
 \ifnum #1\expandafter \@firstoftwo
 \else \expandafter \@secondoftwo
\fi
}%
\providecommand \enquote [1]{``#1''}%
\providecommand \bibnamefont  [1]{#1}%
\providecommand \bibfnamefont [1]{#1}%
\providecommand \citenamefont [1]{#1}%
\providecommand\href[0]{\@sanitize\@href}%
\providecommand\@href[1]{\endgroup\@@startlink{#1}\endgroup\@@href}%
\providecommand\@@href[1]{#1\@@endlink}%
\providecommand \@sanitize [0]{\begingroup\catcode`\&12\catcode`\#12\relax}%
\@ifxundefined \pdfoutput {\@firstoftwo}{%
 \@ifnum{\z@=\pdfoutput}{\@firstoftwo}{\@secondoftwo}%
}{%
 \providecommand\@@startlink[1]{\leavevmode}%
 \providecommand\@@endlink[0]{}%
}{%
 \providecommand\@@startlink[1]{%
  \leavevmode
  \pdfstartlink
   attr{/Border[0 0 1 ]/H/I/C[0 1 1]}%
   user{/Subtype/Link/A<</Type/Action/S/URI/URI(#1)>>}%
  \relax
 }%
 \providecommand\@@endlink[0]{\pdfendlink}%
}%
\providecommand \url  [0]{\begingroup\@sanitize \@url }%
\providecommand \@url [1]{\endgroup\@href {#1}{\urlprefix}}%
\providecommand \urlprefix [0]{URL }%
\providecommand \Eprint[0]{\href }%
\@ifxundefined \urlstyle {%
  \providecommand \doi [1]{doi:\discretionary{}{}{}#1}%
}{%
  \providecommand \doi [0]{doi:\discretionary{}{}{}\begingroup
  \urlstyle{rm}\Url }%
}%
\providecommand \doibase [0]{http://dx.doi.org/}%
\providecommand \Doi[1]{\href{\doibase#1}}%
\providecommand \bibAnnote [3]{%
  \BibitemShut{#1}%
  \begin{quotation}\noindent
    \textsc{Key:}\ #2\\\textsc{Annotation:}\ #3%
  \end{quotation}%
}%
\providecommand \bibAnnoteFile [2]{%
  \IfFileExists{#2}{\bibAnnote {#1} {#2} {\input{#2}}}{}%
}%
\providecommand \typeout [0]{\immediate \write \m@ne }%
\providecommand \selectlanguage [0]{\@gobble}%
\providecommand \bibinfo [0]{\@secondoftwo}%
\providecommand \bibfield [0]{\@secondoftwo}%
\providecommand \translation [1]{[#1]}%
\providecommand \BibitemOpen[0]{}%
\providecommand \bibitemStop [0]{}%
\providecommand \bibitemNoStop [0]{.\EOS\space}%
\providecommand \EOS [0]{\spacefactor3000\relax}%
\providecommand \BibitemShut [1]{\csname bibitem#1\endcsname}%
\bibitem{Lindell_GenDB}%
  \BibitemOpen
  \bibfield{author}{%
  \bibinfo {author} {\bibfnamefont{I.~V.}\ \bibnamefont{Lindell}}, \bibinfo
  {author}   {\bibfnamefont{A.}~\bibnamefont{Sihvola}},\ }%
  \bibfield{journal}{%
  {\bibinfo {journal} {IEEE Transactions on Antennas and Propagation}}\ }%
  \textbf{\bibinfo {volume} {65}},\ \bibinfo {pages} {4656} (\bibinfo {month}
  {September}\ \bibinfo {year} {2017})%
  \bibAnnoteFile{NoStop}{Lindell_GenDB}%
\bibitem{methods}%
  \BibitemOpen
  \bibfield{author}{%
  \bibinfo {author} {\bibfnamefont{I.~V.}\ \bibnamefont{Lindell}},\ }%
  \emph{\bibinfo {title} {Methods for Electromagnetic Field Analysis}}\
  (\bibinfo {publisher} {Oxford University Press and IEEE Press},\ \bibinfo
  {address} {Oxford},\ \bibinfo {year} {1992, 1995})%
  \bibAnnoteFile{NoStop}{methods}%
\bibitem{Shchukin}%
  \BibitemOpen
  \bibfield{author}{%
  \bibinfo {author} {\bibfnamefont{A.~N.}\ \bibnamefont{Shchukin}},\ }%
  \emph{\bibinfo {title} {Propagation of radio waves}}\ (\bibinfo {publisher}
  {Svyazizdat},\ \bibinfo {address} {Moscow},\ \bibinfo {year} {1940})%
  \bibAnnoteFile{NoStop}{Shchukin}%
\bibitem{Leontovich}%
  \BibitemOpen
  \bibfield{author}{%
  \bibinfo {author} {\bibfnamefont{M.~A.}\ \bibnamefont{Leontovich}},\ }%
  \bibfield{journal}{%
  \bibinfo {journal} {Bulletin of Academy of Sciences USSR, Phys.~Ser.}\ }%
  \textbf{\bibinfo {volume} {8}},\ \bibinfo {pages} {16} (\bibinfo {year}
  {1944})%
  \bibAnnoteFile{NoStop}{Leontovich}%
\bibitem{Jackson}%
  \BibitemOpen \bibfield{author}{%
    \bibinfo {author} {\bibfnamefont{J.~D.}\ \bibnamefont{Jackson}},\
  }%
  \emph{\bibinfo {title} {Classical Electronynamics}}\ (\bibinfo
  {publisher} {John Wiley \& Sons, Inc.},\ \bibinfo {address} {New York},\ \bibinfo
  {year} {1999})%
  \bibAnnoteFile{NoStop}{Jackson}%
\bibitem{Glisson}
  \BibitemOpen
  \bibfield{author}{%
  \bibinfo {author} {\bibfnamefont{A.~W.}\ \bibnamefont{Glisson}},\ }%
  \bibfield{journal}{%
  \bibinfo {journal} {Radio Science}\ }%
  \textbf{\bibinfo {volume} {27}},\ \bibinfo {pages} {935} (\bibinfo {month}
  {November--December}\ \bibinfo {year} {1992})%
\bibitem{ele}
  \BibitemOpen
  \bibfield{author}{%
  \bibinfo {author} {\bibfnamefont{M.}\ \bibnamefont{Selvanayagam}}\ and
  \bibinfo {author} {\bibfnamefont{G.~V.}\ \bibnamefont{Eleftheriades}},\ }%
  \bibfield{journal}{%
  \bibinfo {journal} {Physical Review X}\ }%
  \textbf{\bibinfo {volume} {3}},\ \bibinfo {pages} {041011} (\bibinfo {month}
  {November}\ \bibinfo {year} {2013})%
\bibitem{bilo}
  \BibitemOpen
  \bibfield{author}{%
  \bibinfo {author} {\bibfnamefont{S.}\ \bibnamefont{Vellucci}},\ 
  \bibinfo {author} {\bibfnamefont{A.}\ \bibnamefont{Monti}},\ 
  \bibinfo {author} {\bibfnamefont{A.}\ \bibnamefont{Toscano}}\ and 
  \bibinfo {author} {\bibfnamefont{F.}\ \bibnamefont{Bilotti}},\ }%
  \bibfield{journal}{%
  \bibinfo {journal} {Physical Review Applied}\ }%
  \textbf{\bibinfo {volume} {7}},\ \bibinfo {pages} {034032} (\bibinfo {month}
  {March}\ \bibinfo {year} {2017})%
\bibitem{glybo}
  \BibitemOpen
  \bibfield{author}{%
  \bibinfo {author} {\bibfnamefont{S.~B.}\ \bibnamefont{Glybovski}},\ 
  \bibinfo {author} {\bibfnamefont{S.~A.}\ \bibnamefont{Tretyakov}},\ 
  \bibinfo {author} {\bibfnamefont{P.~A.}\ \bibnamefont{Belov}},\ 
  \bibinfo {author} {\bibfnamefont{Y.~S.}\ \bibnamefont{Kivshar}}\ and 
  \bibinfo {author} {\bibfnamefont{C.~R.}\ \bibnamefont{Simovski}},\ }%
  \bibfield{journal}{%
  \bibinfo {journal} {Physics Reports}\ }%
  \textbf{\bibinfo {volume} {634}},\ \bibinfo {pages} {1} (\bibinfo {year} {2016})%
\bibitem{Alu}
  \BibitemOpen
  \bibfield{author}{%
  \bibinfo {author} {\bibfnamefont{A.}\ \bibnamefont{Al\`u}},\ }%
  \bibfield{journal}{%
  \bibinfo {journal} {Physical Review B}\ }%
  \textbf{\bibinfo {volume} {80}},\ \bibinfo {pages} {241115} (\bibinfo {month}
  {December}\ \bibinfo {year} {2009})%
\bibitem{Bohren_Huffman}%
  \BibitemOpen
  \bibfield{author}{%
  \bibinfo {author} {\bibfnamefont{C.~F.}\ \bibnamefont{Bohren}}\ and\ \bibinfo
  {author} {\bibfnamefont{D.~R.}\ \bibnamefont{Huffman}},\ }%
  \emph{\bibinfo {title} {Absorption and Scattering of Light by Small
  Particles}}\ (\bibinfo {publisher} {John Wiley \& Sons, Inc.},\ \bibinfo {address} {New York},\
  \bibinfo {year} {1983})%
  \bibAnnoteFile{NoStop}{Bohren_Huffman}%
\bibitem{Wiscombe}
  \BibitemOpen
  \bibfield{author}{%
  \bibinfo {author} {\bibfnamefont{W.~J.}\ \bibnamefont{Wiscombe}},\ }%
  \bibfield{journal}{%
  \bibinfo {journal} {Applied Optics}\ }%
  \textbf{\bibinfo {volume} {19}},\ \bibinfo {pages} {1505} (\bibinfo {month}
  {May}\ \bibinfo {year} {1980})%
\bibitem{Brillouin}
  \BibitemOpen
  \bibfield{author}{%
  \bibinfo {author} {\bibfnamefont{L.}\ \bibnamefont{Brillouin}},\ }%
  \bibfield{journal}{%
  \bibinfo {journal} {Journal of Applied Physics}\ }%
  \textbf{\bibinfo {volume} {20}},\ \bibinfo {pages} {1110} (\bibinfo {month}
  {November}\ \bibinfo {year} {1949})%
\bibitem{rs}%
  \BibitemOpen
  \bibfield{author}{%
  \bibinfo {author} {\bibfnamefont{H.}\ \bibnamefont{Wall\'en}},\
  \bibinfo {author} {\bibfnamefont{H.}\ \bibnamefont{Kettunen}} and\
  \bibinfo {author} {\bibfnamefont{A.}\ \bibnamefont{Sihvola}},\ }%
 \bibfield{journal}{%
  \bibinfo {journal} {Radio Science}\ }%
  \textbf{\bibinfo {volume} {50}},\ \bibinfo {pages} {18} (\bibinfo {month}
  {January}\ \bibinfo {year} {2015})%
  \bibAnnoteFile{NoStop}{rs}%
\bibitem{ZBO}%
  \BibitemOpen
  \bibfield{author}{%
  \bibinfo {author} {\bibfnamefont{I.~V.}\ \bibnamefont{Lindell}}\,
  \bibinfo {author} {\bibfnamefont{A.~H.}\ \bibnamefont{Sihvola}},\ 
   \bibinfo {author} {\bibfnamefont{P.}\ \bibnamefont{Yl\"a-Oijala}}\ and\
  \bibinfo {author} {\bibfnamefont{H.}\ \bibnamefont{Wall\'en}},\ }%
 \bibfield{journal}{%
  \bibinfo {journal} {IEEE Transactions on Antennas and Propagation}\ }%
  \textbf{\bibinfo {volume} {57}},\ \bibinfo {pages} {2725} (\bibinfo {month}
  {September}\ \bibinfo {year} {2009})%
  \bibAnnoteFile{NoStop}{ZBO}%
\bibitem{eibert}%
  \BibitemOpen
  \bibfield{author}{%
  \bibinfo {author} {\bibnamefont{Ismatullah}}\ and\
  \bibinfo {author} {\bibfnamefont{T.~F.}\ \bibnamefont{Eibert}},\ }%
 \bibfield{journal}{%
  \bibinfo {journal} {IEEE Transactions on Antennas and Propagation}\ }%
  \textbf{\bibinfo {volume} {57}},\ \bibinfo {pages} {2084} (\bibinfo {month}
  {July}\ \bibinfo {year} {2009})%
  \bibAnnoteFile{NoStop}{eibert}%
\bibitem{pasi}%
  \BibitemOpen
  \bibfield{author}{%
  \bibinfo {author} {\bibfnamefont{P.}\ \bibnamefont{Yl\"a-Oijala}},\ 
   \bibinfo {author} {\bibfnamefont{J.}\ \bibnamefont{Lappalainen}}\ and\
  \bibinfo {author} {\bibfnamefont{S.}\ \bibnamefont{J\"arvenp\"a\"a}},\ }%
 \bibfield{journal}{%
  \bibinfo {journal} {IEEE Transactions on Antennas and Propagation}\ }%
  \textbf{\bibinfo {volume} {66}},\ \bibinfo {pages} {487} (\bibinfo {month}
  {January}\ \bibinfo {year} {2018})%
  \bibAnnoteFile{NoStop}{pasi}%
\bibitem{Dimi_josa}%
  \BibitemOpen
  \bibfield{author}{%
  \bibinfo {author} {\bibfnamefont{D.~C.}\ \bibnamefont{Tzarouchis}},\ 
  \bibinfo {author} {\bibfnamefont{P.}\ \bibnamefont{Yl\"a-Oijala}},\ 
   \bibinfo {author} {\bibfnamefont{T.}\ \bibnamefont{Ala-Nissila}}\ and\
  \bibinfo {author} {\bibfnamefont{A.}\ \bibnamefont{Sihvola}},\ }%
 \bibfield{journal}{%
  \bibinfo {journal} {Journal of the Optical Society of America B}\ }%
  \textbf{\bibinfo {volume} {33}},\ \bibinfo {pages} {2626} (\bibinfo {month}
  {December}\ \bibinfo {year} {2016})%
  \bibAnnoteFile{NoStop}{pasi}%
\bibitem{Lindell06a}%
  \BibitemOpen
  \bibfield{author}{%
  \bibinfo {author} {\bibfnamefont{I.~V.}\ \bibnamefont{Lindell}}\ and\
  \bibinfo {author} {\bibfnamefont{A.~H.}\ \bibnamefont{Sihvola}},\ }%
  \bibfield{journal}{%
  \bibinfo {journal} {IEEE Transactions on Antennas and Propagation}\ }%
  \textbf{\bibinfo {volume} {54}},\ \bibinfo {pages} {3669} (\bibinfo {month}
  {December}\ \bibinfo {year} {2006})%
  \bibAnnoteFile{NoStop}{Lindell06a}%
\bibitem{Munk}%
  \BibitemOpen \bibfield{author}{%
    \bibinfo {author} {\bibfnamefont{B.~A.}\ \bibnamefont{Munk}},\ }%
  \emph{\bibinfo {title} {Frequency Selective Surfaces: Theory and
      Design}}\ (\bibinfo {publisher} {John Wiley \& Sons, Inc.},\ \bibinfo {address}
  {New York},\ \bibinfo {year} {2000})%
  \bibAnnoteFile{NoStop}{Munk}%
\end{thebibliography}

\end{document}